\documentclass{aa} 
\usepackage{natbib}
\usepackage{graphicx}
\usepackage{txfonts}
\usepackage{epsfig}
\usepackage{times}
\usepackage{rotating}
\usepackage{float}
\usepackage[caption = false]{subfig}
\usepackage{setspace}
\usepackage{lscape}
\usepackage{epstopdf}
\usepackage{tabularx}
\usepackage{caption}

\DeclareCaptionFormat{cont}{#1 (cont.)#2#3\par}

\begin{document}

\title{Discovery of a Ly$\alpha$ blob photo-ionised by a super-cluster of massive stars associated to a $z=3.49$ galaxy}

\subtitle{}

\authorrunning{S. Zarattini et al.}

\titlerunning{Photo-ionised blob at $z=3.49$}

\author{S. Zarattini\inst{1,2}, J. M. Rodr\'iguez-Espinosa\inst{1,2,3}, C. Mu\~noz-Tu\~n\'on\inst{1,2},  J.M. Mas-Hesse\inst{4}, and P. Arrabal Haro \inst{5}}

\institute{Instituto de Astrof\'isica de Canarias (IAC), E-38205 La Laguna, Spain 
\and Departamento de Astrof\'isica, Universidad de La Laguna, E-38206 La Laguna, Spain
\and Instituto de Astrof\'isica de Andaluc\'ia, E-18008 Granada, Spain
\and Centro de Astrobiolog\'ia, CSIC--INTA, E-28850 Madrid, Spain
\and NSF’s National Optical-Infrared Astronomy Research Laboratory, 950 N. Cherry Ave., Tucson, AZ 85719, USA
}

\date{\today}
\abstract{}
{We report the discovery and characterisation of a Lyman $\alpha$ (Ly$\alpha$) blob close to a galaxy at redshift $z=3.49$. We present our analysis to check whether the companion galaxy could be the source of the ionised photons responsible for the Ly$\alpha$ emission from the blob. }
{We use images obtained from the 10.4 m Gran Telescopio Canarias (GTC) telescope that are part of the Survey of High-z Absorption Red and Dead Sources (SHARDS) project. The blob is only visible in the F551W17 filter, centred around the Ly$\alpha$ line at the redshift of the galaxy. 
We measure the luminosity of the blob with a two-step procedure. First, we describe the radial surface brightness (SB) profile of the galaxy using a S\'ersic function. We then remove this model from the SB profile of the blob and measure the luminosity of the blob alone. We also estimate the Ly$\alpha$ continuum of the galaxy using an Advanced Camera for Surveys (ACS) image from the Hubble Space Telescope (HST) in the filter F606W, that is wider than the SHARDS one and centred at about the same wavelength. In this image the galaxy is visible, but the blob is not detected, since its Ly$\alpha$ emission is diluted in the larger wavelength range of the F606W filter.}
{We find that the Ly$\alpha$ luminosity of the blob is $1.0\times 10^{43}$ erg s$^{-1}$, in agreement with other Ly$\alpha$ blobs reported in the literature. The luminosity of the galaxy in the same filter is $2.9\times 10^{42}$ erg s$^{-1}$. The luminosity within the HST/ACS image, that we use to estimate the Ly$\alpha$ continuum emission, is $L_{cont} = 1.1 \times 10^{43}$ erg s$^{-1}$. With these values we are able to estimate the Ly$\alpha$ equivalent width (EW), that is found to be $111$ \AA\ (rest-frame). This value is in good agreement with the literature and suggests that a super-cluster of massive ($1-2 \times 10^7$ M$_\odot$) and young ($2-4$ Myr)  stars could be responsible for the ionisation of the blob. We also use two other methods to estimate the luminosity of the galaxy and the blob, in order to assess the robustness of our results. We find a reasonable agreement that supports our conclusions. It is worth noting that the Ly$\alpha$ blob is  spatially decoupled from the galaxy by 3 GTC/SHARDS pixels, corresponding to 5.7~kpc at the redshift of the objects. This misalignment could suggest the presence of an ionised cone of material escaping from the galaxy, as found in nearby galaxies such as M82.}
{}


\maketitle

\section{Introduction}
\label{sec:intro}
Lyman $\alpha$ ($Ly\alpha$) blobs (LABs) are enigmatic objects that were discovered about 20 years ago \citep{Fynbo1999,Steidel2000}. They could have been produced either by photo-ionisation, galactic super-winds/outflows, resonant scattering of Ly$\alpha$ photons from starbursts, or active galactic nuclei (AGNs). 

The scenario of the galactic super winds is supported by several observations. In particular, \citet{Wilman2005} suggested that the  Ly$\alpha$ extended emission of a star-forming galaxy at $z=3.09$ was resulting from the submill powered by a burst of stellar formation some $10^8$ years before, combined with cooling radiation. \citet{Matsuda2004} found a star-formation rate of at least 600 M$_\odot$ yr$^{-1}$ for the LAB1 blob, in agreement with submillimeter observations \citep{Chapman2001}. \citet{Ohyama2003} also supported the wind-driven origin of the Ly$\alpha$ emission. In particular, these authors were able to study the kinematic properties of the blob, concluding that the hypothesis of galactic super winds must be preferred over the other scenarios, at least for some cases.

On the other hand, other studies strongly support the AGN-driven mechanism as the responsible of the ionisation. For example, some of the blobs have radio emission that is spatially correlated with the Ly$\alpha$ one \citep{Miley2008}, implying that the AGN may be powering the extended Ly$\alpha$ emission. LAB1, previously considered a LAB powered by an extremely strong starburst \citep{Matsuda2004}, was found by \citet{Overzier2013} to be indeed powered by a strong AGN, and has become, since then, a typical case-study. Indeed, it is argued that, in particular, the most luminous LABs could be associated to AGNs \citep{Geach2009,Kim2020}.

The cooling radiation from cold streams of gas accreting onto galaxies was proposed as an alternative origin \citep{Nilsson2006,Smith2007,Scarlata2009}. However, more recent studies tend to disfavour this scenario \citep{Yang2011,Yang2014,Prescott2015}. In particular, \citet{Prescott2015} showed that their results disfavour the cooling radiation as the main mechanism in the production of the Ly$\alpha$ emission since such process would not be expected to produce strong HII emission as observed. Moreover, the presence of CIV and CIII indicates that the gas is enriched.

\begin{figure*}
    \centering
    \includegraphics[trim=0 0 0 0,width=\textwidth]{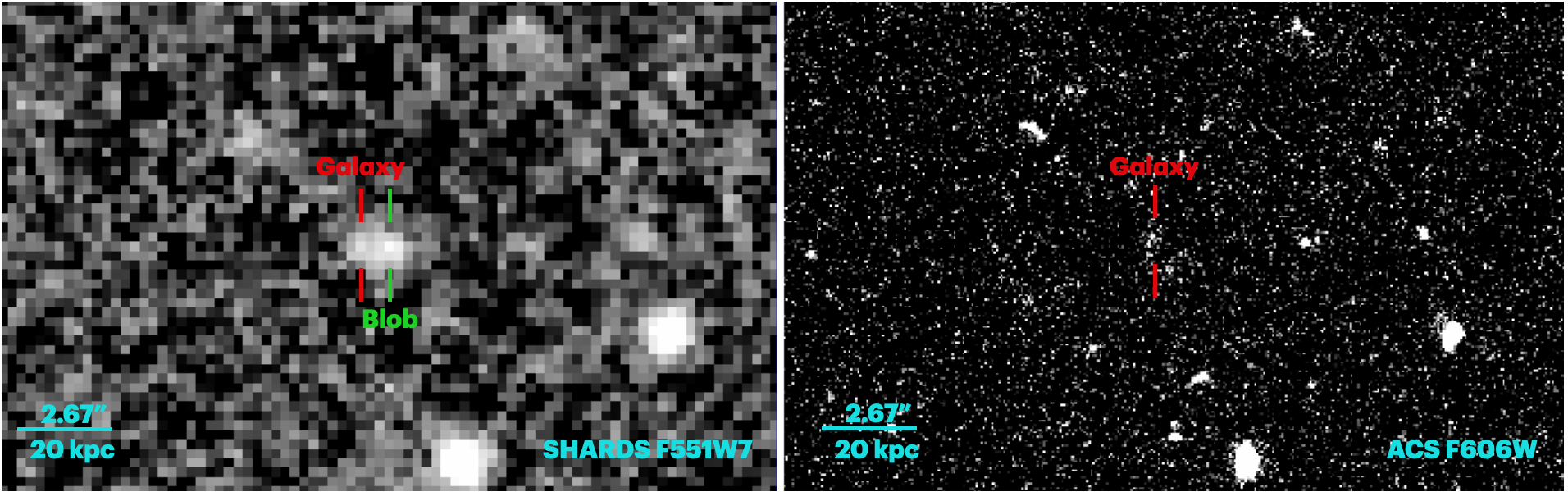}
    \caption{Left panel: the image of the galaxy plus blob in the SHARDS F551W17 filter. Right panel: same region, but in the HST/ACS F606W filter. In the HST/ACS filter the blob is not visible. }
    \label{fig:blob_general}
\end{figure*}

It is currently accepted  that 74\% LABs and 78\% bright LABs are  located in overdense regions, which is in agreement with the trend found in the literature that LABs generally locate in overdense regions \citep[e.g.][and references therein]{Ramakrishnan2023}. LABs, or at least extended  Ly$\alpha$ emission, have been discovered around many quasi-stellar objects (QSOs) and radio-galaxies at high redshift through slit spectroscopy \citep{Heckman1991,Lehnert1998,Bunker2003,Villar-Martin2003,vanBreugel2006A,Willott2011,North2012}, narrow-band imaging \citep{Hu1987, Steidel2000,Smith2007,Yang2014}, and recently, using integral field spectroscopy \citep[hereafter IFS,][]{Christensen2006,Francis2006, Herenz2015, Borisova2016}.

These blobs are thus signposts of intense star formation and their Ly$\alpha$ luminosities usually varies in the range $10^{42}-10^{44}$ erg s$^{-1}$ \citep{Prescott2015,Caminha2016, Kimock2021}. In this work we report the discovery and characterisation of a Ly$\alpha$ blob at $z=3.49$ thanks to the Survey for High-z Absorption Red and Dead Sources (SHARDS) of the Gran Telescopio Canarias (GTC) telescope. 

The cosmology used in this paper is H$_0 = 69.6$, $\Omega_m = 0.286$ and $\Omega_m = 0.714$ \citep{Bennett2014}.

\section{Data}
\label{sec:data}
Most of the data used in this work are taken from the SHARDS survey, an ESO/GTC common large program that explored the GOODS-N field in the wavelength range $5000-9500$ \AA. The program obtained images in 25 contiguous medium-band filters, providing sub-arcsecond seeing, a 3$\sigma$ depths of at least 26.5 AB magnitudes in all bands, and a spectral resolution of $R\sim 50$ \citep{PerezGonzalez2013} in about 200 hours of observations with the OSIRIS instrument of the 10.4 m GTC telescope.
This deep and multi-wavelength dataset enabled the first simultaneous search of Lyman-alpha emitters and Lyman-break galaxies \citep{Arrabal2018}. In their work, those authors presented a list of about 1500 Lyman-alpha emitters (LAEs) and Lyman-break galaxies (LBGs). For most of these sources, \citet{Barro2019} computed the photometric redshift using the photometric information of different telescopes and instruments. In particular, the objects with SHARDS data were included in their Tier 2 sample, for which the available dataset comprise HST optical and infrared images, ultraviolet data from the Kitt Peak and Large Binocular telescopes, and infrared data from a variety of instruments/telescopes, including Subaru and Spitzer. On this extended dataset, \citet{Barro2019} applied a spectral energy distribution (SED) fitting procedure based on the EAZY algorithm \citep{Brammer2008}, thus providing a robust photometric redshift for the Tier 2 sources.

The blob that we study in this work was found by our team in a subsequent exploration of the SHARDS images close to the galaxy SHARDS20018464 and it is only visible in the F551W17 image, that is reported in the left panel of Fig.~\ref{fig:blob_general}. This filter spans the wavelength range $5430-5590$\AA. For comparison, we show in the right panel of the same figure the HST/ACS image in the F606W filter of the same field. The latter spans approximately the wavelength range $4700-7000$, thus including the F551W17 region. The galaxy is visible in both images, but the blob is only visible in the SHARDS one. In all the other filters, a weak emission can be seen eventually in the position of the galaxy, not in that of the blob. The misalignment between the blob and the galaxy is visible only in the F551W17 filter and it's the only candidate in our sample with this unique characteristics, making it an ideal target for studying the mechanism that could be responsible of its ionisation.

The photometric redshift of SHARDS20018464 is $z = 3.49\pm0.06$ \citep{Barro2019,Arrabal2020}. At this redshift, the Ly$\alpha$ emission  (at 1215.67 \AA)  moves to 5458.36 \AA, that is within the SHARDS F551W17 filter range. The fact that SHARDS20018464 has his most-brilliant emission in this filter and that the blob is only visible in the same spectral range and just 3 pixels away from the galaxy position strongly suggests that the blob is linked to the parent galaxy.

\section{Flux measurements}
\label{sec:mesurements}
We run SExtractor \citep{Bertin1996} to characterise the galaxy+blob detection in the SHARDS filter. In the header of the image we found the calibration constant, that is 33.0785 mag in the AB system.

We use this calibration to estimate the AB magnitudes in the rest of our work and we also compute the surface brightness (SB) zero point using the formula $ZP = 33.0785+5\times {\rm log}(0.254)$, where $0\farcs254$ is the pixel scale of the image. This ZP will be used for calibrating the surface brightness profile of the object.  

We report the values of the SExtractor detection in the corresponding section of Table \ref{tab:results}. The AB magnitude of the object is 25.35 mag, estimated using the {\tt mag\_auto} parameter. SExtractor is not able to separate the galaxy from the blob and for this reason we only use this measurement as a reference.

\subsection{Galaxy and blob deconvolution}
\begin{figure}[t]
    \includegraphics[trim=-20 0 40 0,width=0.45\textwidth]{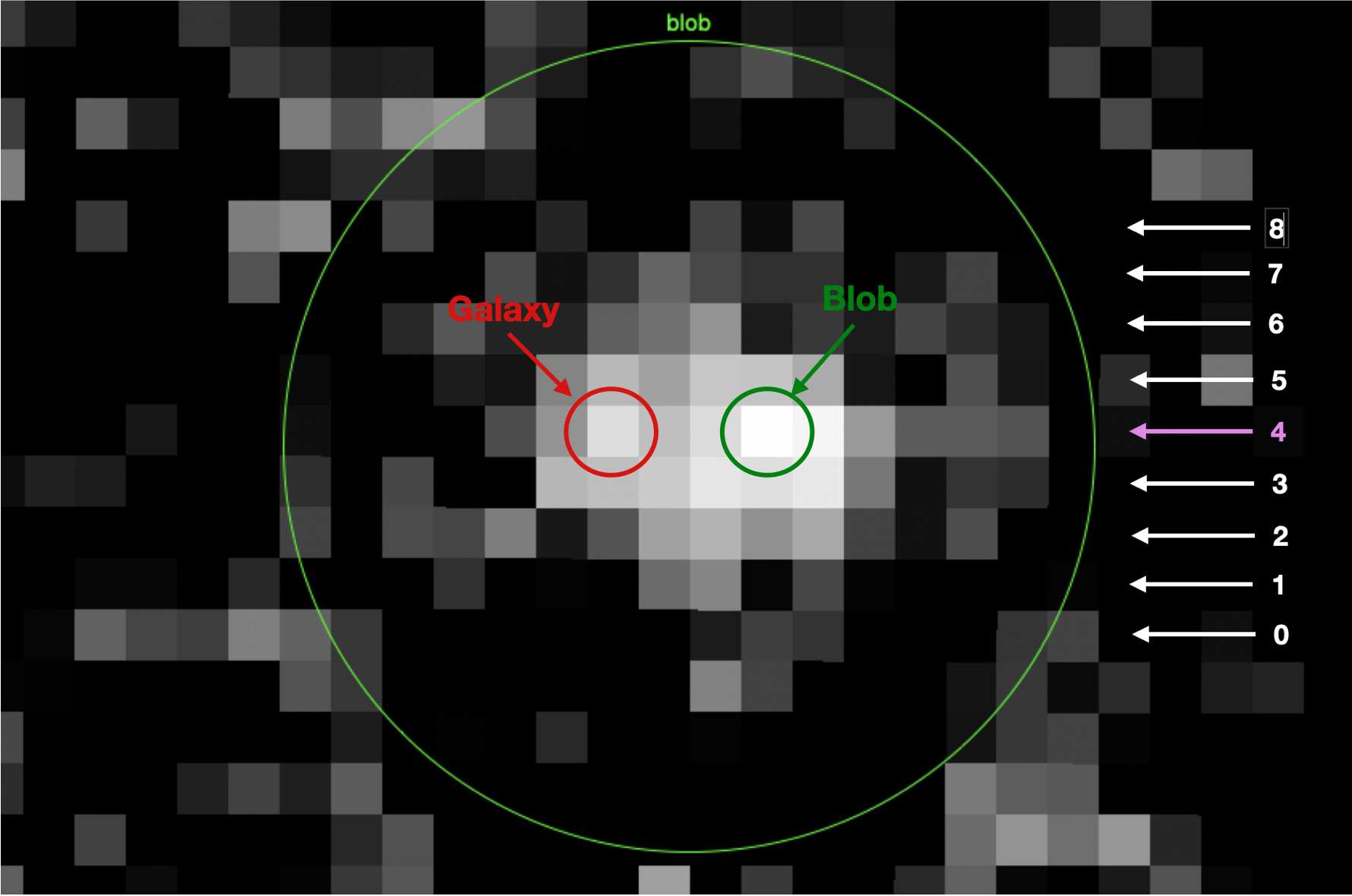}
    \caption{The blob and its host galaxy. The large green circle represents a region of 2\farcs0 around the detected object, whereas the arrows on the right side represent the nine lines that we plotted individually in Fig. \ref{fig:gauss1}, with the one connecting the two centres highlighted in violet. Moreover, the small red and green circles are the centres of the galaxy and the blob, respectively.}
    \label{fig:blob_zoom_positions}
\end{figure}

The main method adopted in this work for measuring the luminosity of the galaxy and the blob is to estimate the excess of light after removing a S\'ersic profile. Since the blob and the galaxy have their centres apparently aligned on the CCD (same Y pixel coordinate), we use the centre of the galaxy (red circle in Fig. \ref{fig:blob_zoom_positions}) as the starting point to analyse the four directions parallels to the axes of the image (from the centre to the top, bottom, left, and right). In this way, we have three directions in which we can assume that only the galaxy contributes and one in which the blob is dominant.

\begin{figure}
\centering
    \includegraphics[trim=20 0 40 100,width=0.5\textwidth]{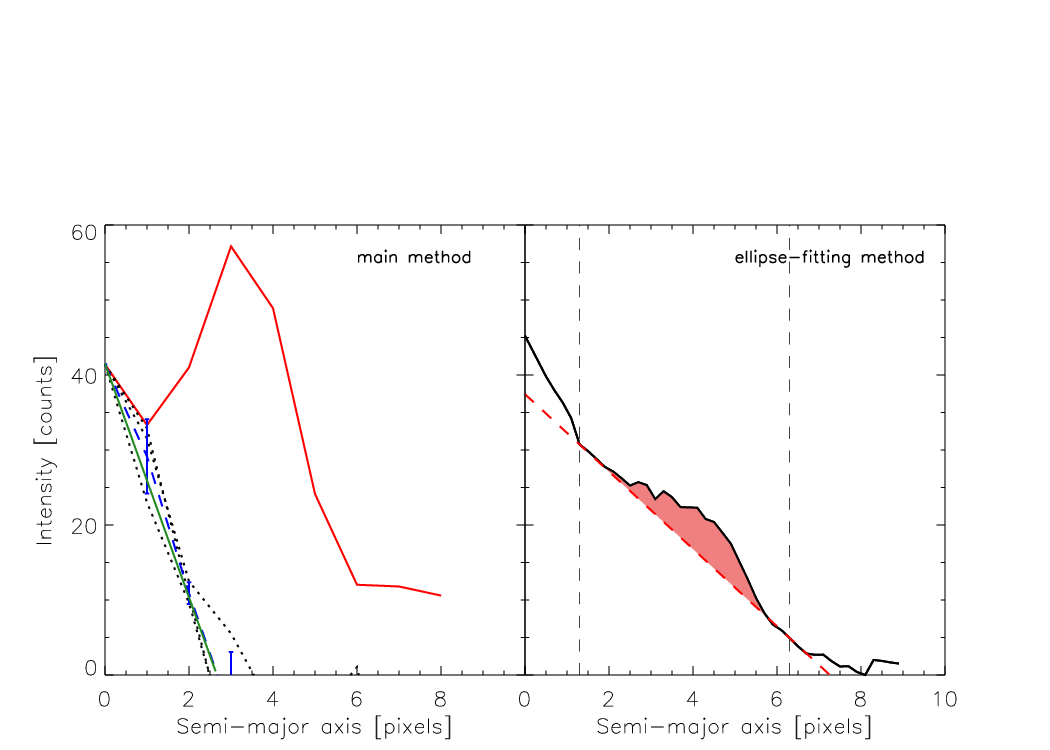}
    \caption{Left panel: radial profiles of the galaxy in the three directions where the blob is not present (black dotted lines). The mean profile is represented with the blue dashed line (with corresponding uncertainties) and the S\'ersic profile that we choose to represent the profile is shown with the green solid line. The red solid line represents the profile of the blob. Right panel: radial dependency of the intensity of the ellipses fitted with the IRAF task {\tt ellipse}. The peak centred at about sma=4 corresponds to the region of the blob. The red dashed line is connecting two points of the profile: the internal one is where there is an abrupt change in the slope of the profile, whereas the external point is the last point above the zero level. These two points are indicated by the two vertical lines. The coral area represent the difference between the profile and the red dashed line.}
    \label{fig:gasp2d}
\end{figure}

The results are shown in Fig. \ref{fig:gasp2d}. Black dotted lines are the three directions where only the galaxy is dominant and indeed they follow a quite similar profile. In blue, we overplotted the mean of these three lines (with the relative error bars, given by the dispersion of the data) and the green solid line is a S\'ersic profile with S\'ersic index $n=1$, effective radius $r_e = 0.38$ and effective intensity $I_e = 7.8$. We note that this profile is not a fit, since the number of available points is too small. However, it follows quite well the blue line and it remains within the errors, so we can consider it as a reasonable representation of the profile of the galaxy.

The red solid line represents the profile in the direction of the blob. It can be seen that, in this kind of decomposition, the blob starts dominating already in the second pixel and its flux peaks at about the third pixel ($\sim 0\farcs8$ or $\sim 5.7$ kpc). 

We then estimate the luminosity of the blob computing the difference between the area under the red solid line and the area of the green solid line that we adopted as the model of the galaxy in this approximation. This is an upper limit to our estimates, since we assume that the blob is the strongest contributor to the emission. However, it is probably the most precise description of what we are observing, since the model of the galaxy is coherent in three out of four directions and we can thus expect that in the fourth direction its contribution remains the same.

The luminosity of the blob is $1.04\times 10^{43}$, the one of the galaxy is $2.87\times 10^{42}$, and the total flux is $1.33\times 10^{43}$. All the results are reported in Table \ref{tab:results}.

 We also estimate the flux with other two methods, to confirm the robustness of the results and to give a lower limit to the Ly$\alpha$ emission from the blob. In the first comparison method we identify the two peaks with the maximum emission in the F551W17 filter. These peaks correspond, in our interpretation, to the centre of the galaxy and the blob, respectively (as in the main method). In Fig. \ref{fig:blob_zoom_positions} both centres are visible along the same row, the blob being that on the right and  highlighted with a small green circle, whereas the galaxy is on the left, highlighted with the small red circle.

\begin{figure}[t]
    \includegraphics[trim=20 20 0 20,width=0.5\textwidth]{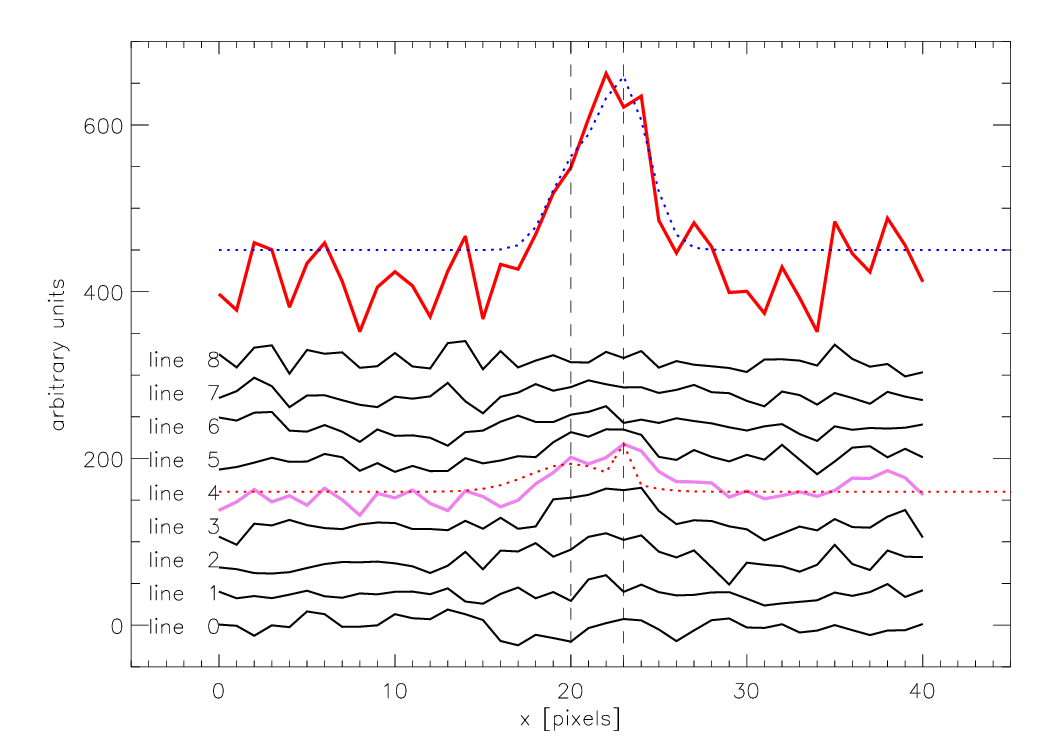}
    \caption{Intensity plots of the 9 rows around the peaks of the blob and its host galaxy. The pixels row containing the emission peaks of the blob and the galaxy companion is highlighted in violet, whereas in red we show the combination of all the 9 lines fitted with a double Gaussian function (blue dotted line). The vertical axis is in arbitrary unit since, for sake of clarity, we add a vertical offset of 40 ADU to every individual row. Moreover, we also add 450 ADUs to the counts of the red solid line.}
    \label{fig:gauss1}
\end{figure}

We thus focus our attention on a tiny region of 9 rows around the position where the peaks of the blob and its host galaxy are found (indicated with arrows in the right side of Fig \ref{fig:blob_zoom_positions}). Each of these rows is represented with the corresponding intensity plot in Fig. \ref{fig:gauss1}. The intensity of these nine lines are shown individually, with the violet one representing the row that includes the two peaks (line 4). Moreover, we show in red the intensity profile of the collapsed lines (e.g. the pixel-to-pixel sum of the nine lines). We then fit a double Gaussian's function to this red row (galaxy+blob), in order to estimate the contribution of each component. Using this method, we estimate a total magnitude of 25.59 AB and a total luminosity of $3.34\times 10^{42}$ erg s$^{-1}$. To do this, we compute the flux in ADUs for the sum of all the pixels of the nine lines and then apply the calibration constant to get the AB magnitude and the flux.

\begin{figure}[t]
    \includegraphics[trim=10 10 0 20,width=0.5\textwidth]{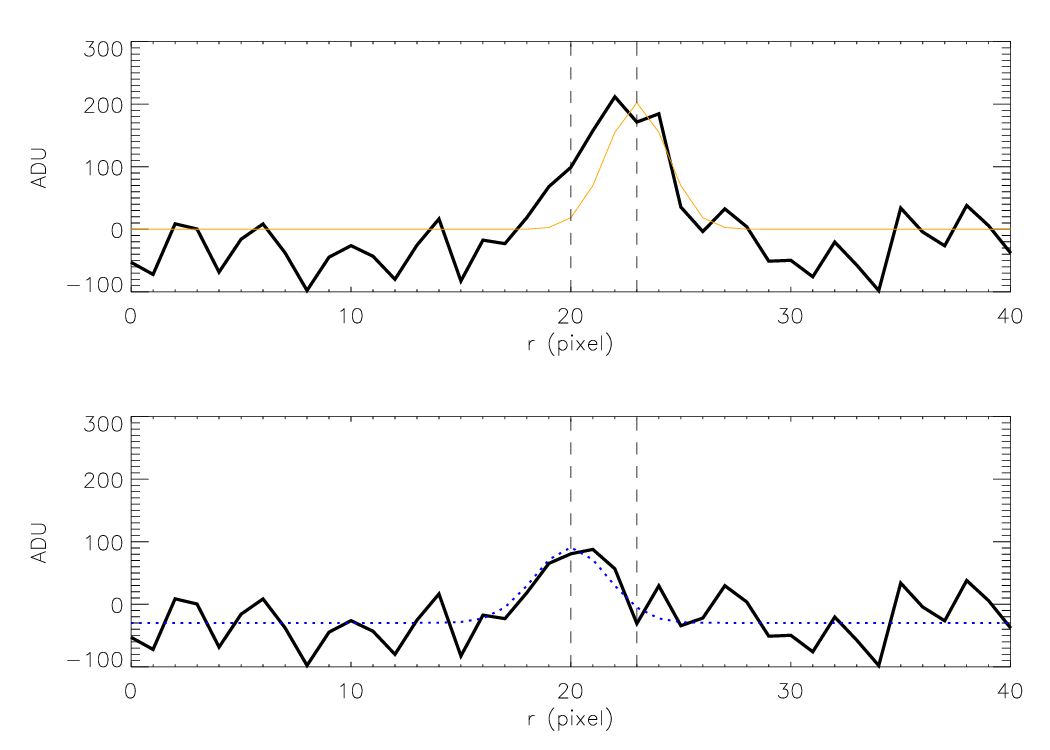}
    \caption{Intensity plot of the combination of the 9 lines around the galaxy/blob. In the top panel, the black line is the same as the red line of Fig. \ref{fig:gauss1}) and the peak corresponding to the blob is fitted with a single Gaussian function. In the bottom panel, such Gaussian is removed from the profile and a different Gaussian function is fitted to the remaining signal, corresponding to the flux of the host galaxy. The vertical dashed lines are the same as in Fig. \ref{fig:gauss1} and represent the peaks of the emission of the galaxy and the blob.}
    \label{fig:gauss2}
\end{figure}

We can also estimate the flux in ADUs (and magnitudes) for the Gaussian of the blob and the galaxy separately in the following way: first, we fit a single Gaussian to the position of the blob (thus forcing the peak in a specific position, see top panel of Fig. \ref{fig:gauss2}). We then remove this Gaussian from the profile and we fit another one to the residuals, that in this case represents the underlying galaxy (bottom panel of Fig. \ref{fig:gauss2}). For both Gaussians we can estimate the ADU fluxes and the AB magnitudes of the two components as in the previous paragraph. The results are reported in Table \ref{tab:results}.

The Ly$\alpha$ luminosities of the blob and the galaxy, according to this method and assuming that all the light in the F551W17 filter is due to the Ly$\alpha$ emission, are $2.34\times 10^{42}$ and $2.40\times 10^{42}$ erg s$^{-1}$, respectively. The total luminosity  is $3.34\times 10^{42}$ erg s$^{-1}$, since in this approximation the object is slightly fainter than when compared to SExtractor results. This can be due to the different estimation of the sky level, since for the Gaussian fitted to the blob it is assumed that the sky starts at zero ADUs, whereas it is probably fainter.

Another comparison method is based on the isophotes fitting of the galaxy using the IRAF\footnote{IRAF is distributed by the National Optical Astronomy Observatories, which are operated by the Association of Universities for Research in Astronomy, Inc., under cooperative agreement with the National Science Foundation.} task {\tt ellipse}. The centre of the fit is the peak of the galaxy (e.g. the pixel highlighted with the small red circle in Fig. \ref{fig:blob_zoom_positions}) and the ellipse fitting was performed using fixed semi-major axis progression, with step of 0.2 pixels. We did not mask the image, since there is no visible contamination from external sources and we need to maximise the number of pixels available for the analysis. From {\tt ellipse}, we obtain the intensity profile, shown in the right panel of Fig. \ref{fig:gasp2d} with the black thick solid line. 

We then try to perform the photometric decomposition of the galaxy using {\tt GASP2D} \citep{Mendez-abreu2008}, but we were not able to find a good fit, probably due to the fact that the code was written to work with large galaxies (more than 100 pixel in radius) and in this case the number of available pixels is too small $( < 10)$. For this reason we construct the line connecting two peculiar points of the profile, shown as a dashed-red line in the right panel of Fig. \ref{fig:gasp2d}. The inner one is the point (at 1.3 pixel) at which the slope of the profile changes abruptly (e.g. as if the galaxy was dominated by a bulge in the inner part and a disc in the external regions). The external point is the last point of the ellipse fitting where the ellipse is fitted and not just a copy of the previous one (6.3 pixel). It is worth noting that this is just an approximation, with the idea to have a lower limit to the luminosity of the blob. That is, we are assuming that the larger fraction of the light has to be associated with the galaxy. As a consequence, the peak of the blob is diluted. Moreover, since the {\tt ellipse} procedure is providing the radial profile of the entire ellipse, each point of the blob is also diluted in all the other points of the isophote, mostly dominated by the companion galaxy.

We estimate the flux of the blob as the difference between the area under the black thick curve and the one under the red dashed line in the selected region (e.g. included within the two vertical lines at 1.3 and 6.3 pixels in Fig. \ref{fig:gasp2d}). In the same figure, we show this area in coral. The red dashed line represents our model of the galaxy in that region. 
The total magnitude of the galaxy (without the contribution of the blob) is obtained by adding the area under the red dashed line in the range $1.3-6.3$ pixels to the area of the central region (e.g. the area under the black curve with radius smaller than 1.3 pixels). The result is a magnitude estimated between $0-6.3$ pixels.

From the area of the blob, we estimate a magnitude of 26.35 mag, fainter than the previous estimation of 25.97 that we got with our main method. The flux of the blob is $1.65\times 10^{42}$ erg s$^{-1}$ and indeed we stress that, with this method, we are giving a lower limit to the luminosity of the blob, assuming that most of the light is due to the companion galaxy. As a consequence, the magnitude of the galaxy is 24.90 mag with this method and the total one is 24.61 mag. The latter estimation is thus brighter than the SExtractor one.

\begin{table*}[]
        \caption{Properties of SHARDS20018464 and of its two components.}
	\begin{center}
    \begin{tabular}{cccccccc}
    \hline\hline
          ID & R.A. & Dec.  & L & L$_\lambda$ & EW (Ly$\alpha$) & EW$_{0}$(Ly$\alpha$) & F$_\nu$ \\
          &  &  & [erg s$^{-1}$] & [erg s$^{-1} \AA^{-1}$] & \AA & \AA  & [AB mag] \\
         \hline
         \multicolumn{8}{c}{\rule{0pt}{4ex} SExtractor global properties (galaxy+blob)} \\
         \hline
         SHARDS20018464 & 189.28& 62.23  & $3.5 \times 10^{42}$ & -  & - & - & 25.35 \\
         \hline
         \multicolumn{8}{c}{\rule{0pt}{4ex} First method: excess of light in the SB profile after removing a S\'ersic profile} \\
         \hline
         blob (Ly$\alpha$) & - & - &   $1.04 \times 10^{43}$ & - & - & - &24.35 \\
         gal (continuum) & - & - &  $2.87 \times 10^{42}$ & $2.08 \times 10^{40}$ & 500 & 111 & 25.75 \\
         tot & - & - &  $1.33 \times 10^{43}$ & - & - & - &24.09 \\\hline
         \multicolumn{8}{c}{\rule{0pt}{4ex} Second method: double Gaussian fit} \\
         \hline
         blob (Ly$\alpha$) & -  & -  & $2.34 \times 10^{42}$ & - & - & - & 25.97 \\
         gal (continuum) & -  & -   & $2.40 \times 10^{42}$ & $1.74 \times 10^{40}$ & 134 & 30 & 25.94 \\
         tot & -  & -   & $3.34 \times 10^{42}$ & - & - & - & 25.59 \\
         \hline
         \multicolumn{8}{c}{\rule{0pt}{4ex} Third method: excess of light in the SB profile} \\
         \hline
         blob (Ly$\alpha$) & - & -  & $1.65 \times 10^{42}$ & - & - & - & 26.35 \\
         gal (continuum) & - & - &  $6.26 \times 10^{42}$ & 4.53 $\times 10^{40}$ & 36 & 8 & 24.90 \\
         tot & - & - &  $8.17 \times 10^{42}$ & - & - & - & 24.61 \\
         \hline
         \multicolumn{8}{c}{\rule{0pt}{4ex} UV continuum estimated using the F606W HST/ACS filter} \\
         \hline
         HST/ACS F606W & 189.28 & 62.23 &  $1.08 \times 10^{43}$ & $5.78 \times 10^{39}$ &  - & - & 26.99 \\
         \hline
         \multicolumn{8}{c}{\rule{0pt}{4ex} Combined SHARDS F551W17 and HST/ACS F606W  filters  } \\
         \hline
         Ly$\alpha$ &-&-& 2.7 $\times 10^{42}$ &  $5.78 \times 10^{39}$ & 467 & 104 & -\\ 
         \hline
    \end{tabular}
    \end{center}
    \tablefoot{First line: properties of the SHARDS20018464 object (galaxy+blob) as obtained with SExtractor on the SHARDS F551W17 filter. The magnitude is calibrated using the zero point given by the SHARDS team.  The values per wavelength units correspond to the effective centre wavelength of the filter. Second/third/fourth lines: our main method, based on modelling the galaxy with a S\'ersic profile and then computing the area of the blob as the difference between the total one and the area of the modelled galaxy. Fifth/sixth/seventh lines: values from the areas of the Gaussians. The total one is the area of the double Gaussian, the other are the individual ones. Eighth/ninth/tenth lines: values from the ellipse fitting method assuming that the galaxy is the dominant component. The total flux iso is the same as in the previous method, we computed those of the blob and the galaxy by applying the fraction of the area that corresponds to each of them (e.g the blob is about 65\% of the area, so 65\% of the flux iso).  The last two lines show the photometry with the broad band HST/ACS F606W filter and the values of $L(Ly\alpha$) and $EW(Ly\alpha$) derived combining the SHARDS narrow band filter and the HST/ACS broadband one. The  $L(Ly\alpha$) value is computed subtracting from the total luminosity in the SHARDS filter the proportional continuum contribution from the HST/ACS F606W filter.}
    \label{tab:results}
\end{table*}

\subsection{Morphology and sizes}

It is worth noticing that in the SHARDS filters the object has a very small angular size, so that its apparent size on the SHARDS image is dominated by the seeing. In our specific image, the seeing was computed with IRAF task {\tt imexam} as 1\farcs0, corresponding to 3.9 pixels. As shown in Fig.~\ref{fig:gasp2d}, the separation of the galaxy and blob peaks is only $\sim 3$ pixels, just resolved under the seeing conditions. The pixel scale of our observations is 0\farcs254 per pixel, so that the separation is 0\farcs76. Since the object is found at $z=3.49$, the physical scale is 7.478 kpc/\arcsec. Therefore, the physical separation between the galaxy and the blob is only 5.7 kpc. 

\subsection{Continuum}

In order to properly estimate the continuum luminosity we use the HST image available from the CANDELS program \citep{Koekemoer2011}. We use the HST/ACS F606W broadband filter, since its range is 4700-7000 \AA, thus including the F551W17 SHARDS filter.
We run SExtractor on the image as we did for the SHARDS one and repeat the same steps in order to estimate the luminosity of the galaxy in this band. The results are reported in Table \ref{tab:results}. We also include the value of the continuum per unit wavelength at the filters centre (5809.26~\AA, corresponding to 1293~\AA\ rest-frame for the redshift of the object), in erg s$^{-1}$ $\AA^{-1}$. 

SExtractor also estimates the Kron radius of the galaxy, that is 5.79 pixels. The scale of the ACS camera is 0\farcs05, thus the Kron radius of the galaxy in the ACS F606W broadband filter is 2.2 kpc. This size is about 50\% of the size of the galaxy in the SHARDS filter, thus confirming that seeing is a dominant effect in those images. However, the blob is only visible in the narrower SHARDS filter, which reinforces the idea of it being a blob of emission produced by the Ly$\alpha$ photons in a very narrow wavelength range.

\subsection{ ${Ly\alpha}$ equivalent width}

We measure the ${Ly\alpha}$ equivalent width (EW) associated to the different deconvolutions of the galaxy (continuum) and the blob (mostly ${Ly\alpha}$ emission), as well as directly combining the SHARDS F551W17 luminosity with the continuum level estimated from the  HST/ACS F606W broadband filter. We consider the latter value ($EW({Ly\alpha}) = 104 $ \AA\ rest-frame) as the most reliable one, since it is model independent. We want to stress that the equivalent width so derived is completely consistent with the results of the deconvolution by the S\'ersic profiles ($EW({Ly\alpha}) = 111 $ \AA\ rest-frame). These values are also consistent with the Ly$\alpha$ EW estimated for this source in \citet{Arrabal2018} solely based on the SHARDS observations (126  \AA). The derived $EW({Ly\alpha})$ and $L({Ly\alpha})$ values for all the methods are listed in Table \ref{tab:results}. 

\section{Discussion and conclusions}
The illustration that is, in our opinion, best describing the blob is the one that we presented as our main method, that is the excess of light in the SB profile after removing a S\'ersic model of the galaxy. In this section, we are going to discuss mainly this result. However, we stress that it is compatible with the other two methods, which remain useful to assess the quality of the result and to provide a lower limit to the emission of the blob.

The morphology of this LAB is similar to several LAB encountered by \citet{Matsuda2011} in their survey of blobs at $z \sim 3$. However, our blob is smaller in size, being at most $5-8$ kpc versus the more than 100 kpc of all the blobs in their sample.

The host galaxy was modelled with a S\'ersic profile with $n=1$, meaning that this is a disc galaxy, in agreement with expectations from \citet{Kartaltepe2023}, who found that 60\% of galaxies at $z=3$ are disc galaxies. The size of the galaxy is small, with a Kron radius of 2.2 kpc in the HST image. These values are in agreement with \citet{Lumbreras-Calle2019}. In particular, these authors found that star-forming galaxies can be split into two classes, one with small S\'ersic indexes and small effective radii and the other with large S\'ersic indexes and large effective radii. The former are expected to be blue, whereas the latter are red. In a forthcoming work we will analyse the colour of the host galaxy for this object and for a larger sample of $Ly\alpha$ blob candidates, but according to the size and $n$ index we expect that SHARDS20018464 should be a blue galaxy.

As shown in Fig.~\ref{fig:blob_zoom_positions}, the ${Ly\alpha}$ blob is clearly associated to a strong source of UV continuum, though each component peaks at a different location, not being co-spatial. We will first analyse whether the cluster of massive, potentially young stars that originates the UV continuum could also emit enough photo-ionising photons to produce the observed ${Ly\alpha}$ emission. As shown in Table~\ref{tab:results} the global ${Ly\alpha}$ equivalent width of the system should be 111 \AA\ (rest-frame). This value is perfectly consistent with the predictions for strong episodes of massive star formation, as shown by \citet{Rodriguez-Espinosa2021}. Synthesis models predict intrinsic values of ${EW(Ly\alpha)}$ around 100 \AA\ for coeval starbursts at an age around 3~Myr, but they would be also consistent with an extended episode of star formation already in its equilibrium phase (after around 30~Myr), when the number of massive stars that are born balances the ones that finish their lifetime.  

We want to stress that the observed values of ${EW(Ly\alpha)}$ should be considered just as a lower limit, since the ${Ly\alpha}$ photons are strongly affected by scattering and associated destruction by dust, so that their escape fraction is usually well below $f_{esc,Ly\alpha} = 1.0$. We have applied the semi-empirical calibration by \citet{Sobral2019} ($f_{esc,Ly\alpha} = 0.0048\times EW(Ly\alpha) \pm 0.05$) to derive an estimate of the ${Ly\alpha}$ escape fraction, resulting in $f_{esc,Ly\alpha} \sim 0.53 \pm 0.05$, which is a relatively large value for a ${Ly\alpha}$ emitting galaxy \citep{Hayes2011}. Correcting the observed ${Ly\alpha}$ emission by this escape fraction, we would get values of the intrinsic ${EW(Ly\alpha)}$ around $208 \pm 20$~\AA, which would bring the age of the starbursts to $\sim 2.5$ Myr for a coeval burst, and around 4~Myr for a more extended episode. Such a large value of the intrinsic ${EW(Ly\alpha)}$ would be incompatible with the formation of massive stars for longer than around 5~Myr. Indeed, any value of  $f_{esc,Ly\alpha} \lesssim 0.9$ would lead to an intrinsic  ${EW(Ly\alpha)} \gtrsim 125$ \AA, too large for a long--lasting star formation episode \citep{Rodriguez-Espinosa2021}.

Synthesis models also allow us to estimate the strength of the star formation process that has originated this ${Ly\alpha}$ blob. Using the calibrations by \citet{Oti-Floranes2010} we find that enough ionising photons would be emitted to yield the observed Ly${\alpha}$ by a coeval star formation episode having transformed $1.2\times 10^7$ $M_\sun$ of gas into stars, as well as by an extended episode at an average star formation rate of 2~$M_\sun$/yr (in both cases normalised to a Salpeter initial mass function in the mass range $2-120$~$M_\sun$). Assuming $f_{esc,Ly\alpha} \sim 0.5$, the values would be $1.7\times 10^7$ $M_\sun$ and 4.6~$M_\sun$/yr, respectively. 

Therefore, we conclude that a massive ($1-2\times 10^7 M_\sun$) super-cluster formed by young (around $2-4$ Myr old) stars would be able to simultaneously produce the observed (rest-frame) ultraviolet continuum and to produce enough ionising photons as required to feed the observed Ly$\alpha$ luminosity. 

\begin{acknowledgements}
The authors thanks the anonymous referee for its useful comments that help clarifying the results of this paper. The authors also thanks Dr. Rosa Calvi for useful comments and discussion. This work is part of the collaboration ESTALLIDOS, supported by the Spanish research grants, PID2019-107408GB-C43 and PID2022-136598NB-C31 and by the Government of the Canary Islands through EU FEDER funding project PID2021010077. S.Z. is also supported by the Ministry of Science and Innovation of Spain, project PID2020-119342GB-I00. Based on observations made with the Gran Telescopio Canarias (GTC), installed at the Spanish Observatorio del Roque de los Muchachos of the Instituto de Astrofísica de Canarias, on the island of La Palma. This work is based on data obtained with the SHARDS filter set, purchased by Universidad Complutense de Madrid (UCM). SHARDS was funded by the Spanish Government through grant AYA2012-31277, and on observations made with the NASA/ESA Hubble Space Telescope obtained from the Space Telescope Science Institute, which is operated by the Association of Universities for Research in Astronomy, Inc., under NASA contract NAS 5–26555.

\end{acknowledgements}

\bibliography{bibliografia}

\end{document}